\documentclass[aps,prb,twocolumn,superscriptaddress,floatfix]{revtex4}
\usepackage{psfig}
\usepackage{graphicx}
\usepackage{epsfig}
\bibstyle{apsrev.bib}
\begin{document}
\input epsf.sty

\title{Nonequilibrium phase transitions and finite size scaling\\ in weighted scale-free
  networks} 
\author{M\'arton Karsai}
\affiliation{Institute of Theoretical Physics, Szeged University, 
H-6720 Szeged, Hungary}
\author{R\'obert Juh\'asz}
\affiliation{Theoretische Physik, Universit\"at des Saarlandes,
D-66041 Saarbr\"ucken, Germany}
\author{Ferenc Igl\'oi}
\affiliation{Research Institute for Solid State Physics and Optics, 
H-1525 Budapest, P.O.Box 49, Hungary}
\affiliation{Institute of Theoretical Physics, Szeged University,
H-6720 Szeged, Hungary}
\date{\today}

\begin{abstract}
  
  We consider nonequilibrium phase transitions in weighted scale-free
  networks, in which highly connected nodes, which are created earlier
  in time are partially immunized. For epidemic spreading we solve the
  dynamical mean-field equations and discuss finite-size scaling
  theory. The theoretical predictions are confronted with the results
  of large scale Monte Carlo simulations on the weighted
  Barab\'asi-Albert network. Local scaling exponents are found
  different at a typical site and at a node with very large
  connectivity.

\end{abstract}


\maketitle

\newcommand{\bc}{\begin{center}}
\newcommand{\ec}{\end{center}}
\newcommand{\be}{\begin{equation}}
\newcommand{\ee}{\end{equation}}
\newcommand{\ba}{\begin{array}}
\newcommand{\ea}{\end{array}}
\newcommand{\beqn}{\begin{eqnarray}}
\newcommand{\eeqn}{\end{eqnarray}}

\section{Introduction}

Complex networks, which have a more complicated topology than periodic
lattices have been observed in a large class of systems in different
fields of science, technics, transport, social and political life,
etc, see Refs.\cite{AB01,DM01,DM03} for recent reviews.  The structure
of complex networks is described by graphs\cite{bollobas}, in which
the nodes represent the agents and the edges the possible
interactions. Generally there are connections between remote sites,
too, which is known as the small world effect\cite{WS98}.  Another
feature of many real networks is the non-democratic way of the
distribution of the links: there are sites which are much more
connected than the average and the distribution of the number of
edges, $P(k)$, has a power-law tail:
\be
P_D(k) \simeq A k^{-\gamma},\quad k \gg 1\;,
\label{gamma}
\ee
thus the edge distribution is scale free. In real networks the degree
exponent is generally: $2<\gamma<3$ and as shown by Barab\'asi and
Albert\cite{BA99} scale-free networks are usually the results of
growth and preferential attachment.

Since the agents of a network interact in one or another way it is
natural to ask about the cooperative behavior of the system. In
particular if there exist same kind of (thermodynamical) phases and if
there are singular points as the strength of the interaction or other
suitable parameter (such as the strength of disordering field,
temperature, etc.) are varied. In this respect static models\cite{stauffer,ising,sajat,potts} (Ising,
Potts models, etc.), as well as non-equilibrium processes\cite{percolation,epidemic,newman}
(percolation, spread of epidemics, etc.) are investigated.  Generally
non-weighted networks are considered, in which the strength of
interaction at each bond is constant. Due to long-range interactions
conventional mean-field behavior is expected to hold, at least if the
network is sufficiently weakly connected, i.e. $\gamma>\gamma_u$.  For
the Ising model $\gamma_u=5$, whereas for the percolation and epidemic
spreading $\gamma_u=4$.  At $\gamma=\gamma_u$ there are logarithmic
corrections to the mean-field singularities, whereas for
$\gamma_u>\gamma>\gamma_c$ we arrive to the unconventional mean-field
regime in which the critical exponents are $\gamma$ dependent. The
effect of topology of scale-free networks becomes dramatic for $\gamma
\le \gamma_c$, when the average of $k^2$, $\langle k^2 \rangle$, as
well as the strength of the average interaction becomes divergent.
Consequently for any finite value of the interaction scale-free
networks are in the ordered state, c.f. there is no threshold value of
epidemic spreading. Since $\gamma_c=3$, in realistic networks with
homogeneous interactions always this type of phenomena should occur.

Recently, much attention has been paid on weighted networks, in which
the interactions are not homogeneous. Generally the strength of
interactions of highly connected sites are comparatively weaker than
the average, which can be explained by technical or geographical
limitations.  To model epidemic spreading one should keep in mind that
sites with a large coordination number are generally earlier connected
to the network and in the long period of existence they have larger
chance to be immunized.

An interesting class of degraded networks has been introduced recently
by Giuraniuc et al\cite{joseph} in which the strength of interaction
in a link between sites $i$ and $j$ is rescaled as:
\be
\lambda_{i,j}=\lambda \frac{(k_i k_j)^{-\mu}}{<k^{-\mu}>^2}\;,
\label{lambda}
\ee
where $k_i$ and $k_j$ are the connectivities in the given sites. The
properties of this type of network in equilibrium critical phenomena,
in particular for the Ising model have been studied in detail in
Ref\cite{joseph}. Most interestingly the equilibrium critical behavior
is found to depend on the effective degree exponent,
\be
\gamma'=(\gamma-\mu)/(1-\mu)\;,
\label{gamma'}
\ee
thus topology and interaction seem to be converted. One important
aspect of the degraded network in Eq.(\ref{lambda}) that phase
transition in realistic networks with $\gamma \le 3$ is also possible,
if the degradation exponent, $\mu$, is sufficiently large. Therefore
theoretical predictions about critical singularities can be confronted
with the results of numerical calculations.

In this paper we study nonequilibrium phase transitions in weighted
networks. Our aim with these investigations is twofold. First, we want
to check if the simple reparametrization rule in Eq.(\ref{gamma'})
stays valid for nonequilibrium phase transitions, too. For this
purpose we make dynamical mean-field calculations and perform large
scale Monte Carlo simulations. Our second aim is to study the form of
finite-size scaling in nonequilibrium phase transitions in weighted
scale-free networks. To analyze our numerical results we use recent
field-theoretical calculations\cite{lubeck} in which finite-size
scaling in Euclidean lattices above the upper critical dimension has
been studied. In the conventional mean-field regime of scale-free
networks analogous scaling relations are expected to apply.

The structure of the paper is the following. The dynamical mean-field
solution of the problem is presented in Sec.\ref{sec:mf}, whereas
finite-size scaling theory is shown in Sec.\ref{sec:fss}.  Results of
Monte Carlo simulations of the contact process on weighted
Barab\'asi-Albert networks are presented in Sec.\ref{sec:mc} and
discussed in Sec.\ref{sec:disc}.

\section{Dynamical mean-field solution}
\label{sec:mf}

In the calculation we consider the contact process\cite{harris}, which is the
prototype of a non-equilibrium phase transition in the directed
percolation universality class\cite{Grass}. In this process site $i$ of the
network is either vacant ($\emptyset$) or occupied by at most one
particle ($A$). The dynamics of the model is given by a continuous
time Markov process and is therefore defined in terms of transition
rates. The reactions in the system are of two types: a) branching in
which particles are created at empty sites (provided one of its
neighbors, $j$, is occupied) occurs with rate $\lambda_{i,j}$, b) death of
particles with a rate, $\mu$. This latter rate we set $\mu=1$.
In this section we solve the problem in the mean-field approximation,
which is expected to be exact, due to long-range interactions in the system.

We start with the set of equations for the time
derivative of the mean-value of the density, $\rho_i$, at site,
$i=1,2,\dots,N$:
\be 
\frac{\partial \rho_i}{\partial t}= \sum_j \lambda_{i,j}(1-\rho_i) \rho_j - \rho_i\;,
\label{mf}
\ee
and correlations in the densities at different sites are omitted.
In the next step in the spirit of the mean-field approach we replace
the interactions:
\be
\lambda_{i,j}=\lambda \frac{k_i k_j}{\sum_j k_j}\frac{(k_i k_j)^{-\mu}}{<k^{-\mu}>^2}\;,
\label{lambda_mf}
\ee
i.e. there is an interaction between each site the (mean) value of
which is proportional to the probability of the existence of that
bond. Now in terms of an average density:
\be
\rho=\frac{\sum_j k_j^{1-\mu} \rho_j}{\sum_j k_j^{1-\mu}}\;,
\label{rho}
\ee
the dynamical mean-field equations in Eq.(\ref{mf}) are given by:
\be 
\frac{\partial \rho_i}{\partial t}= \tilde{\lambda} k_i^{1-\mu} (1-\rho_i) \rho - \rho_i\;,
\label{mf1}
\ee
with $\tilde{\lambda}=\lambda <k^{1-\mu}>/<k><k^{-\mu}>^2$.
In the stationary state, $\partial \rho_i/\partial t=0$, the local densities
are given by:
\be \rho_i=\frac{\tilde{\lambda} k_i^{1-\mu}
  \rho}{1+\tilde{\lambda} k_i^{1-\mu} \rho}\;,
\label{rho_i}
\ee
i.e. they are proportional to $k_i^{1-\mu}$. Putting $\rho_i$ from
Eq.(\ref{rho_i}) into Eq.(\ref{rho}) we obtain an equation for $\rho$:
\be
<k^{1-\mu}>=\tilde{\lambda} \int_{k_{min}}^{k_{max}} P_D(k) \frac{k^{2(1-\mu)}}
{1+\tilde{\lambda} k_i^{1-\mu} \rho} {\rm d} k \;,
\label{rho1}
\ee
where summation over $i$ is replaced by an integration over the degree
distribution, $P_D(k)$, and in the thermodynamic limit the upper limit
of the integration is $k_{max} \to \infty$.  The solution of
Eq(\ref{rho1}) in the vicinity of the transition point, $\rho \ll 1$,
depends on the large-$k$ limit of the degree distribution in
Eq.(\ref{gamma}) and is given in terms of the integration variable,
$k'=k^{1-\mu}$, as:
\be
<k'>=\tilde{\lambda} A \int_{k'_{min}}^{k'_{max}} k'^{-\gamma'} \frac{k'^{2}}
{1+\tilde{\lambda} k' \rho} {\rm d} k' \equiv Q(\rho,\gamma'),\quad \rho \ll 1 \;,
\label{rho2}
\ee
where $\gamma'$ is defined in Eq.(\ref{gamma'}). Note, that the
functional form of the equation in (\ref{rho2}) is identical to that
for standard scale-free networks, just with an effective degree
exponent, $\gamma'$. Thus the solution in Ref.\cite{epidemic} can be
applied and in this way we have obtained an extension of the results
in Ref.\cite{joseph} for nonequilibrium phase transitions.

To analyze the solution of Eq.(\ref{rho2}) we apply the method in
Ref.\cite{sajat}, which is somewhat different from the original method
in Ref.\cite{epidemic}.

\begin{itemize}

\item $\gamma' >4$
  
  For small $\rho$, $Q(\rho,\gamma')$ in Eq.(\ref{rho2}) can be
  expanded in a Taylor series at least up to a term with $\sim \rho^2$.
  Consequently there is a finite transition point,
  $\tilde{\lambda}_c=<k'>/A<k'^2>$, and the density in the vicinity of
  the transition point behaves as: $\rho(\lambda) \sim
  (\lambda_c-\lambda)$. This is the conventional mean-field regime. At the borderline case,
$\gamma'=4$, there are logarithmic corrections to the mean-field singularities.

\item $3 < \gamma' < 4$
  
  For small $\rho$ only the linear term in the Taylor expansion of
  $Q(\rho,\gamma')$ exists. The $\rho$-dependence of the next term,
  $a_2(\rho)$, is singular and given by:
\be
a_2(\rho)=-\tilde{\lambda}^2 \rho  A \int_{k'_{min}}^{k'_{max}} k'^{-\gamma'} \frac{k'^{3}}
{1+\tilde{\lambda} k' \rho} {\rm d} k'\;.
\label{a2}
\ee
The $\rho$-dependence can be estimated by noting that for a small, but
finite $\rho$ there is a cut-off value, $\tilde{k'} \sim 1/\rho$, so
that
\be
a_2(\rho) \sim -\tilde{\lambda}^2 \rho  A \int_{k'_{min}}^{\tilde{k'}} k'^{-\gamma'} k'^{3} {\rm d} k'
\sim \rho^{\gamma'-3}\;.
\label{a22}
\ee
Consequently the density at the transition point behaves anomalously,
\be
\rho \sim (\lambda_c-\lambda)^{\beta},\quad \beta=1/(\gamma'-3)\;.
\label{unconv}
\ee
This is the unconventional mean-field region.

\item $\gamma' < 3$
  
  In this case $Q(\rho,\gamma)$ is divergent for small $\rho$. Its
  behavior can be estimated as in Eq.(\ref{a22}) leading to
  $Q(\rho,\gamma') \sim \lambda^{\gamma'-2} \rho^{\gamma'-3}$. Consequently the
  system for any non-zero value of $\lambda$ is in the active phase.
  As $\lambda$ goes to zero the density vanishes as:
\be
\rho \sim \lambda^{(\gamma'-2)/(3-\gamma')}\;.
\label{ord}
\ee
Here at the border, $\gamma'=3$, the system is still in the active phase, but the
density is related to a small $\lambda$ as: $|\ln (\rho\lambda)| \sim 1/\lambda$.
\end{itemize}

Before we confront these analytical predictions with the results of numerical
simulations we discuss the form of finite-size scaling in scale-free networks.

\section{Finite-size scaling}
\label{sec:fss}

In a numerical calculation, such as in Monte Carlo (MC) simulations,
one generally considers systems of finite extent and the properties of
the critical singularities are often deduced via finite-size scaling.
It is known in the phenomenological theory of equilibrium critical
phenomena that due to the finite size of the system, $L$, critical
singularities are rounded and their position is shifted\cite{barber}. As it is elaborated
for Euclidean lattices finite-size
scaling theory has different forms below and above the upper critical
dimension, $d_c$. For $d<d_c$ in the scaling regime the singularities
are expected to depend on the ratio, $L/\xi$, where $\xi$ is the
spatial correlation length in the infinite system\cite{fisherbarber}. On the other hand
for $d > d_c$, when mean-field theory provides exact values of the
critical exponents, finite-size scaling theory involves dangerous
irrelevant scaling variables\cite{fisher}, which results in the breakdown of
hyperscaling relations. For equilibrium critical phenomena predictions
of finite-size scaling theory\cite{brezin} above $d_c$ are checked numerically,
but the agreement is still not satisfactory\cite{numerics}.

For non-equilibrium critical phenomena finite-size scaling above $d_c$
has been studied only very recently\cite{lubeck} and here we recapitulate
the main findings of the analysis. For directed percolation,
which represents a broad class of universality\cite{hinrichsen}, dangerous
irrelevant scaling variables are identified in the fixed point.
As a consequence scaling of the order-parameter is anomalous:
\be
\rho=L^{-\beta/\nu^*} \tilde{\rho}(\delta L^{1/\nu^*}, h L^{\Delta/\nu^*})\;,
\label{fss}
\ee
Here, $\delta$, is the reduced control parameter, with the notations of
Sec. \ref{sec:mf} $\delta=(\lambda-\lambda_c)/\lambda_c$ and $h$ is the
strength of an ordering field. The critical exponents, $\beta=1$ and
$\Delta=2$, are the same as in conventional mean-field theory. The
finite-size scaling exponent is given by, $\nu^*=2/d$, and thus
depends on the spatial dimension, $d$. Note, that below $d_c=4$ it is
the correlation length exponent, $\nu$, which enters into the scaling
expression in Eq.(\ref{fss}), but above $d_c$, due to dangerous
irrelevant scaling variables it should be replaced by $\nu^*$. At the
critical point, $\delta=0$, the scaling function, $\tilde{\rho}(0,x)$, has
been analytically calculated and checked by numerical calculations.

In the following we translate the previous results for complex network, in
which finite-size
scaling is naturally related to the volume of the network, which is
given by the number of sites, $N$. In the conventional mean-field
regime with the correspondence, $N \leftrightarrow L^d$, we arrive
from Eq.(\ref{fss}) to the finite-size scaling prediction:
\be
\rho_{typ}=N^{-\beta/2} \tilde{\rho}_{typ}(\delta N^{1/2}, h N^{\Delta/2})\;,
\label{fss1}
\ee
which is expected to hold for a typical site, i.e. with a
coordination number, $k \sim \langle k \rangle$. On the other hand for
the maximally connected site with $k_{max} \sim N^{1/(\gamma-1)}$
according to Eq.(\ref{rho_i}) the finite-size scaling form is modified by:
\be
\rho_{max}=N^{-\beta/2+(1-\mu)/(\gamma-1)} \tilde{\rho}_{max}(\delta N^{1/2}, h N^{\Delta/2})\;.
\label{fss_max}
\ee
Since in the derivation of the relation in Eq.(\ref{fss}) the actual
value of $\beta$ has not been used, we conjecture that the results in
Eqs.(\ref{fss1}) and (\ref{fss_max}) remain valid in the
unconventional mean-field region, too.

%
%

\section{Monte Carlo simulation}
\label{sec:mc}
In the actual calculation we considered the contact process on the Barab\'asi-Albert scale-free
network\cite{BA99}, which has a degree exponent, $\gamma=3$, and used a degradation
exponent, $\mu=1/2$. Consequently from Eq.(\ref{gamma'}) the effective
degree exponent is $\gamma'=5$, thus conventional mean-field behavior
is expected to hold. (We note that the same system is used to study
the equilibrium phase transition of the Ising model in
Ref.\cite{joseph}.) Networks of sites up to $N=1024$ are generated by
starting with $m_0=1$ node and having an average degree: $\langle k \rangle=2$.
Results are averaged over typically $10000$ independent realizations
of the networks.

In the calculation we started with a single particle at site $i$,
(which was either a typical site or the maximally connected site) and
let the process evolve until a stationary state is reached in which
averages become time independent.  In particular we monitored the
average value of the occupation number, $\rho_i$, as introduced in
mean-field theory in Eq.(\ref{mf}), and the fraction of occupied
sites, $m_i$, (order parameter) as a function of $\lambda$, whereas
$\mu$ was set to be unity.  In the stationary state and in the
vicinity of the transition point $\rho_i$ and $m_i$ are expected to be
proportional with each other and characterize the order in the
system\cite{m_i}.

\begin{figure}
  \begin{center}
     \includegraphics[width=3.35in,angle=0]{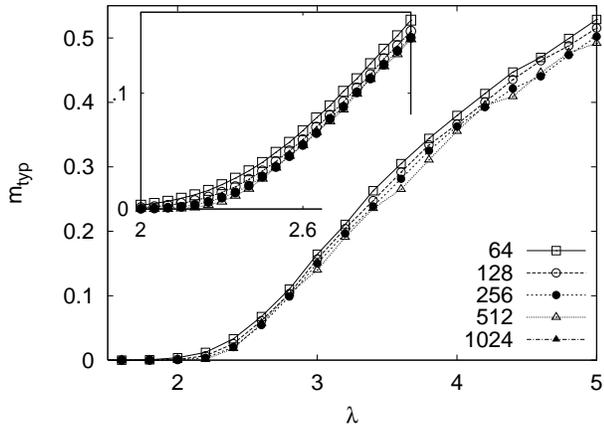}
   \end{center}
   \caption{ Variation of the order parameter at a typical site as a function of the
creation rate, $\lambda$. Inset: Enlargement in the critical region.}
   \label{fig1}
 \end{figure}

\begin{figure}
  \begin{center}
     \includegraphics[width=3.35in,angle=0]{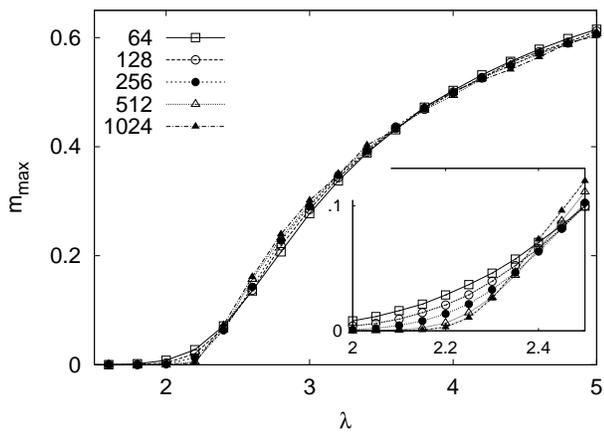}
   \end{center}
   \caption{ As in Fig.\ref{fig1} for the maximally connected site.}
   \label{fig2}
 \end{figure}

The $\lambda$ dependence of the order parameter is shown in
Fig.\ref{fig1} for a typical site and in Fig.\ref{fig2} for the
maximally connected site. Evidently there is a phase transition in the
system in the thermodynamic limit, which is rounded by finite size
effects as shown in the insets of Figs. \ref{fig1} and \ref{fig2}.

\begin{figure}
  \begin{center}
     \includegraphics[width=3.35in,angle=0]{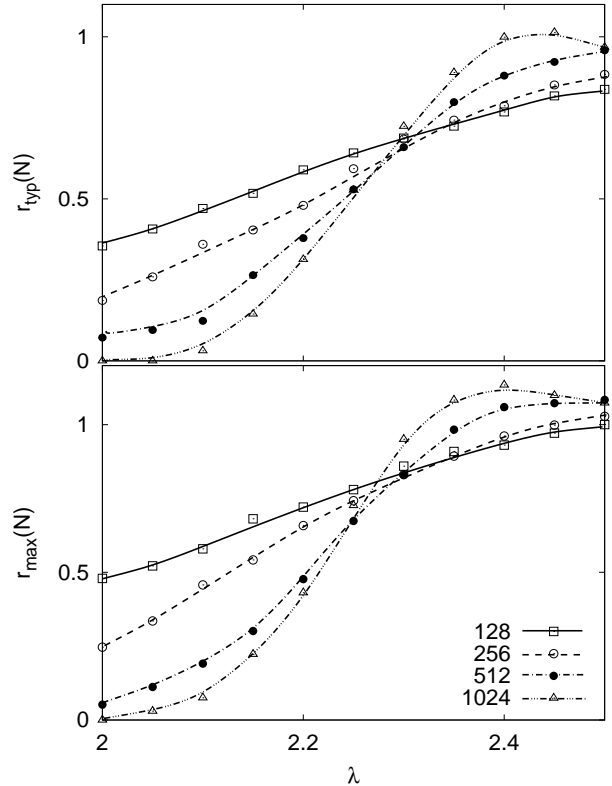}
   \end{center}
   \caption{ The ratio $r(N)=m(N)/m(N/2)$ a) for a typical site, b) for the maximally connected site.
Note that the location of the crossing points, which defines $\lambda_c$ is the same in the two
cases, whereas the value at crossing, which is related to the finite-size exponent, $x$, through
$r(N,\lambda_c)=2^{-x}$ is different.}
   \label{fig3}
 \end{figure}

 To locate the phase transition point we form the ratios:
 $r(N)=m(N)/m(N/2)$, for different finite sizes. As shown in
 Fig.\ref{fig3} $r(N)$ tends to zero in the inactive phase, $\lambda <
 \lambda_c$, and tends to a value of one in the active phase, $\lambda
 > \lambda_c$. The curves for different $N$ cross each other and the
 crossing point can be used to identify $\lambda_c$ through
 extrapolation. Furthermore the value of the ratio at the critical
 point is given by: $r(N,\lambda_c)=2^{-x}$, where $x$ is the
 finite-size scaling exponent, as given in scaling theory in
 Eqs.(\ref{fss1}) and (\ref{fss_max}).

 The transition point is found to be the same within the error of the
 calculation both for a typical site and for the maximally connected
 site and given by: $\lambda_c=2.30(1)$. The finite-size scaling
 exponent, however, calculated from $r(N,\lambda_c)$ is different in
 the two cases. For a typical site we estimate: $x_{typ}=0.54(5)$,
 which should be compared with the field-theoretical prediction in
 Eq.(\ref{fss1}), which is $x_{typ}=\beta/2=1/2$. In the maximally
 connected site the finite-size scaling exponent is measured as
 $x_{max}=0.27(3)$, which again agrees well with the field-theoretical
 prediction in Eq.(\ref{fss_max}):
 $x_{max}=\beta/2-(1-\mu)/(\gamma-1)=1/4$.

\begin{figure}
  \begin{center}
     \includegraphics[width=3.35in,angle=0]{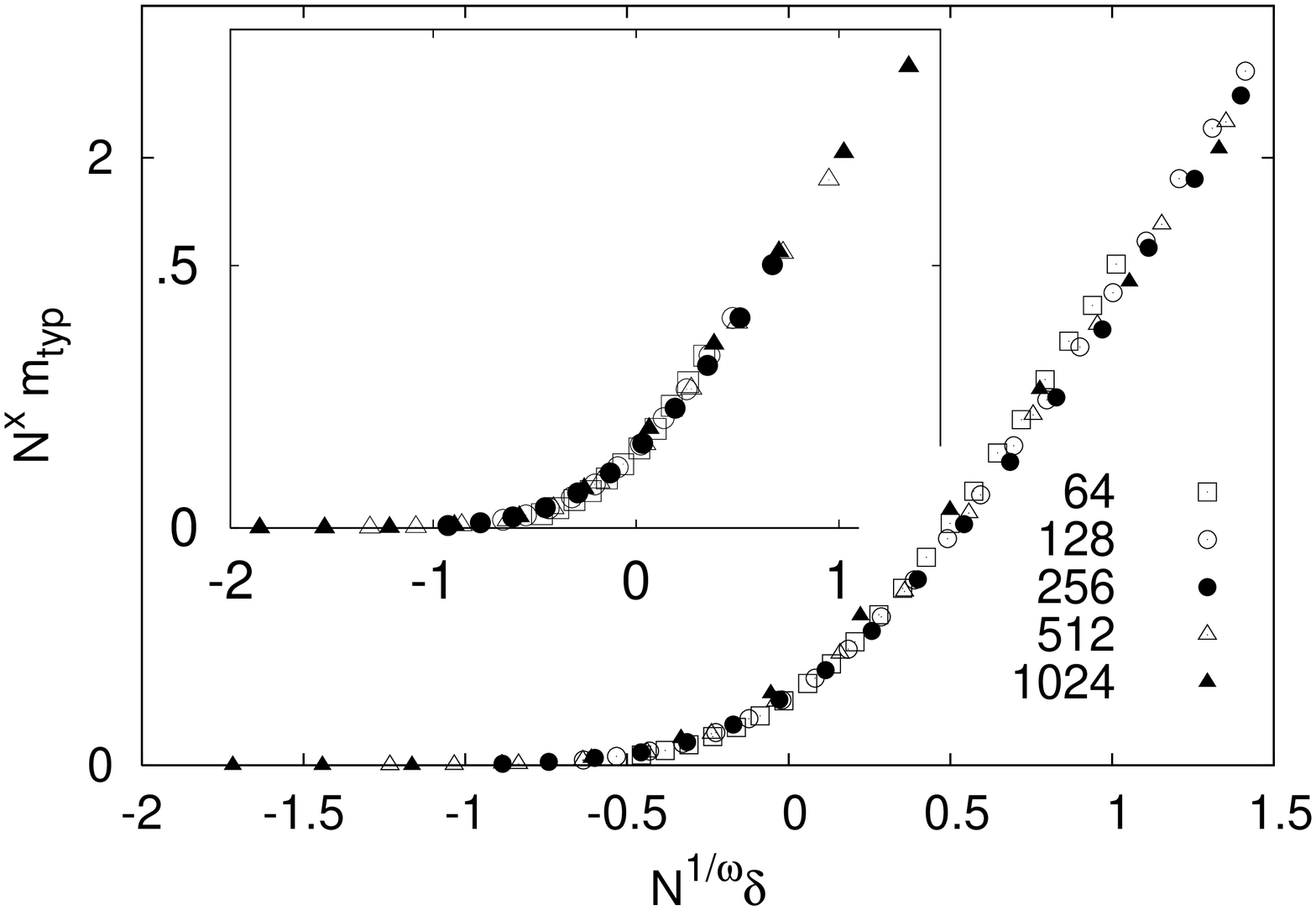}
   \end{center}
   \caption{ Scaling collapse of the order-parameter near the transition point using the
functional forms in Eq.(\ref{fss1}) for a typical site (a) and in Eq.(\ref{fss_max}) for the
maximally connected site (b). The finite-size scaling exponents, $x_{typ}$ and $x_{max}$ are
fixed by the previous analysis and the correlation exponent, $\omega$, is used to have
an optimal collapse of the data, see text.}
   \label{fig4}
 \end{figure}

Next, we consider correlations in the vicinity of the transition point
and calculate the relation between the correlated volume, $\cal{V}$, and the
distance from the critical point, $\delta$, which is expected to be in a
power-law form, $\cal{V} \sim |\delta|^{-\omega}$. According to field-theoretical
results in Eqs.(\ref{fss1}) and (\ref{fss_max}) this exponent is
$\omega=2$, both at a typical site and at the maximally connected
site. Now in the limiting case, ${\cal V} \sim N$, the scaled order-parameter,
$\tilde m= m N^{x}$, is expected to depend on the scaling combination, $N^{1/\omega} \delta$,
which is demonstrated in Fig.
\ref{fig4}, both in the typical site (a) and in the maximally
connected site (b). In both cases $x$ and $\lambda_c$ are fixed by the
previous analysis and $\omega$ is obtained from the optimal scaling
collapse, as $\omega_{typ}=2.05(10)$ and $\omega_{max}=2.00(5)$. Thus, once
more we have a good agreement with the field-theoretical results.

\begin{figure}
  \begin{center}
     \includegraphics[width=3.35in,angle=0]{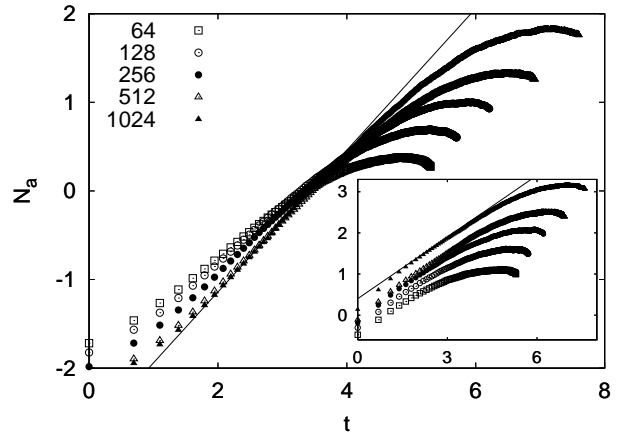}
   \end{center}
   \caption{Time dependence of the number of active sites starting from a single seed
(a) a typical site, b) the maximally connected site) at the critical point in a log-log scale,
Due to finite-size effects the curves are saturated for large time. The slope of the
straight part of the curves defines the effective exponent, $a_{eff}$,
which has strong finite-size dependence. We draw the staight lines for the largest size having
a) $a_{eff}=.8$ and b) $a_{eff}=.5$.}
   \label{fig5}
 \end{figure}

 Finally, we turn to analyze dynamical scaling in the system. At the
 critical point, $\lambda=\lambda_c$ we have measured the number of
 active sites, $N_a$, as a function of time, $t$, which is shown in
 Fig.\ref{fig5} in a log-log plot, when the starting point is a
 typical site or the maximally connected site.  As seen in
 Fig.\ref{fig5} this relation is asymptotically given by $N_a \sim
 t^a$, where the effective value of the critical exponent, $a$, has a
 strong size dependence, in particular by starting with a typical
 site. By extrapolation we obtained $a_{typ}=0.98(5)$, which is
 compatible with the mean-field and finite-size scaling prediction,
 $a_{typ}=1$, whereas starting from the maximally connected site we
 extrapolated $a_{max}=0.57(2)$. Now, keeping in mind that the
 correlated volume can be expressed as, ${\cal V} \sim N \sim N_a/m_i
 \sim N_a^{1/(1-x)}$, we obtain the relation $t \sim {\cal
   V}^{\zeta}$, with $\zeta=(1-x)/a$. For a typical site our numerical
 result, $\zeta_{typ}=0.47(5)$ is compatible with the theoretical
 prediction of $\zeta_{typ}=0.5$, whereas for the maximally connected
 site we obtained $\zeta_{max}=1.28(5)$. From these results, using $N
 \sim \delta^{-\omega}$, we obtain for the scaling behavior of the
 relaxation time, $\tau$, in the vicinity of the transition point,
 $\tau \sim \delta^{-\nu^{\perp}}$, with $\nu^{\perp}=\zeta \omega$,
 so that $\nu^{\perp}_{typ}=0.96(5)$ and $\nu^{\perp}_{max}=2.5(1)$.
 Note that once again at the typical site we are in complete agreement
 with the mean-field and finite-size scaling result,
 $\nu^{\perp}_{typ}=1$.

\section{Discussion}
\label{sec:disc}

We considered non-equilibrium phase trasnsitions in weighted
scale-free networks, in which the creation rate of particles at given
sites is rescaled with a power of the connectivity number. In this way
non-equilibrium phase transitions are realized even in realistic
networks having a degree exponent, $\gamma \le 3$. Mean-field theory,
which is generally belived to be exact in these lattices, is solved
and the previously known three regimes of criticality (conventional
and unconventional mean-field behavior, as well as only active phase)
are identified.  The theoretical predictions in the conventional
mean-field regime are confronted with the results of Monte-Carlo
simulations of the contact process on the weighted Barab\'asi-Albert
network. To analyze the simulation results we have applied and
generalized recent field-theoretical results\cite{lubeck} about
finite-size scaling of non-equilibrium phase transitions above the
upper critical dimension, i.e. in the mean-field regime. For a network
the natural variable is the volume (mass) of the system which enters
in a simple way into the scaling combinations. We have obtained
overall agreement with this finite-size scaling theory in which the
critical exponents are simple rational numbers. We have also
numerically demonstrated that at sites with very large connectivity
there are new local scaling exponents, which differ from the values
measured at a typical site.

F.I. is indebted to S. L\"ubeck for useful discussions and for sending
a copy of Ref.[\onlinecite{lubeck}] before publication.  This work has been
supported by a German-Hungarian exchange program (DAAD-M\"OB), by the
Hungarian National Research Fund under grant No OTKA TO34138, TO37323,
TO48721, MO45596 and M36803. MK thanks to the Institute of Theoretical
Physics, University of Saarland and to Prof.  H. Rieger for
hospitality during an Erasmus exchange.  RJ acknowledges support by
the Deutsche Forschungsgemeinschaft under Grant No.  SA864/2-1.


\begin{thebibliography}{}
\bibitem{m_i} Indeed at the critical point the fraction of active
  sites vanishes with $N$, see in Eqs.(\ref{fss1}) and
  (\ref{fss_max}), therefore starting from a typical site the
  probability of the infection of a largely connected site is
  vanishing.
\end{thebibliography}
\end{document}